\documentclass[cits]{PoS}

\title{Rare kaon decays (NA48/2 and NA62)}

\ShortTitle{Rare kaon decays}

\author{\speaker{Evgueni Goudzovski}\thanks{for the NA48/2 and NA62 collaborations:
F.~Ambrosino, A.~Antonelli, G.~Anzivino, R.~Arcidiacono,
W.~Baldini, S.~Balev, J.R.~Batley, M.~Behler, S.~Bifani, C.~Biino
% ...
}\\
        School of Physics and Astronomy, University of Birmingham, United Kingdom\\
        E-mail: \email{eg@hep.ph.bham.ac.uk}}

\abstract{The rare decay $K^\pm\to\pi^\pm\gamma\gamma$ has been recently measured from data samples collected by the NA48/2 and NA62 experiments at CERN. These measurements are presented, including model-independent spectrum measurements and fits to the Chiral Perturbation Theory description.}

\FullConference{The European Physical Society Conference on High Energy Physics -EPS-HEP2013\\
		18-24 July 2013\\
		Stockholm, Sweden}

\begin{document}

\section{Introduction}

The NA48/2 experiment at the CERN SPS has collected a large sample of charged kaon decays in 2003--04 (corresponding to about $2\times 10^{11}$ $K^\pm$ decays in the vacuum decay volume). The experiment featured simultaneous $K^+$ and $K^-$ beams and was optimized for the search for direct CP violating charge asymmetries in the $K^\pm\to3\pi$ decays~\cite{ba07}. Its successor, the NA62-$R_K$ experiment, collected a 10 times smaller $K^\pm$ decay sample with low intensity beams and minimum bias trigger conditions in 2007--08. NA62-$R_K$ used the same detector as NA48/2, while the data taking conditions were optimized for a measurement of the ratio of the rates of the $K^\pm\to\ell^\pm\nu$ decays ($\ell=e,\mu$)~\cite{la13}. In particular, the main trigger chain required the presence of an electron ($e^\pm$).

The large data samples accumulated by both experiments have allowed precision studies of a range of rare $K^\pm$ decay modes. Recent measurements of the rare decay $K^\pm\to\pi^\pm\gamma\gamma$ (denoted $K_{\pi\gamma\gamma}$ below) from the above data samples are reported here. The NA48/2 and NA62 results on semi-leptonic and leptonic decays of the charged kaon, which have also been presented at this conference, are discussed elsewhere in these proceedings~\cite{ke4, riccardo}.

\section{Beam and detector}

The beam line has been designed to deliver simultaneous narrow momentum band $K^+$ and $K^-$ beams derived from the primary 400 GeV/$c$ protons extracted from the CERN SPS. Secondary beams with central momenta of 60 GeV/$c$ (for NA48/2) or 74 GeV/$c$ (for NA62-$R_K$) were used. The beam kaons decayed in a fiducial decay volume contained in a 114~m long cylindrical vacuum tank. The momenta of charged decay products were measured in a magnetic spectrometer, housed in a tank filled with helium placed after the decay volume. The spectrometer comprised four drift chambers (DCHs), two upstream and two downstream of a dipole magnet which provided a horizontal transverse momentum kick of $120~\mathrm{MeV}/c$ (for NA48/2) or $265~\mathrm{MeV}/c$ (for NA62-$R_K$) to singly-charged particles. Each DCH was composed of eight planes of sense wires. A plastic scintillator hodoscope (HOD) producing fast trigger signals and providing precise time measurements of charged particles was placed after the spectrometer. Further downstream was a liquid krypton electromagnetic calorimeter (LKr), an almost homogeneous ionization chamber with an active volume of 7 m$^3$ of liquid krypton, $27X_0$ deep, segmented transversally into 13248 projective $\sim\!2\!\times\!2$~cm$^2$ cells and with no longitudinal segmentation. The LKr information is used for photon measurements and charged particle identification. An iron/scintillator hadronic calorimeter and muon detectors, not used in the discussed analysis, were located further downstream. A detailed description of the detector can be found in Ref.~\cite{fa07}.

\boldmath
\section{The $K^\pm\to\pi^\pm\gamma\gamma$ decay in the Chiral Perturbation Theory}
\unboldmath

Measurements of radiative non-leptonic kaon decays provide crucial tests of Chiral Perturbation Theory (ChPT) describing weak low energy processes. The
$K_{\pi\gamma\gamma}$ decay has attracted the attention of theorists over the last 40 years~\cite{se72, ec88, da96, ge05}, but remains among the least experimentally studied kaon decays.

\begin{figure}[tb]
\begin{center}
\resizebox{0.45\textwidth}{!}{\includegraphics{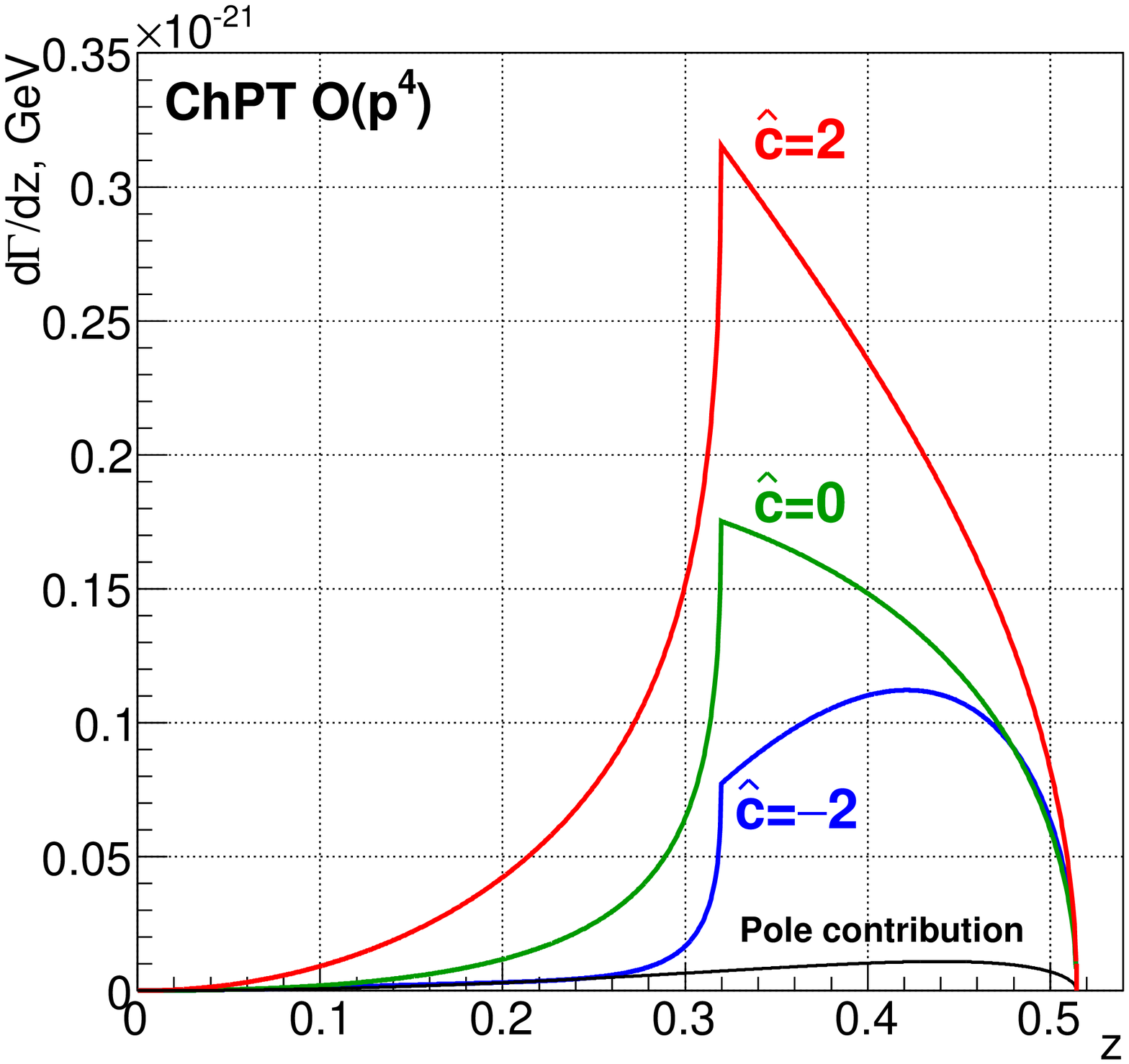}}%
\resizebox{0.45\textwidth}{!}{\includegraphics{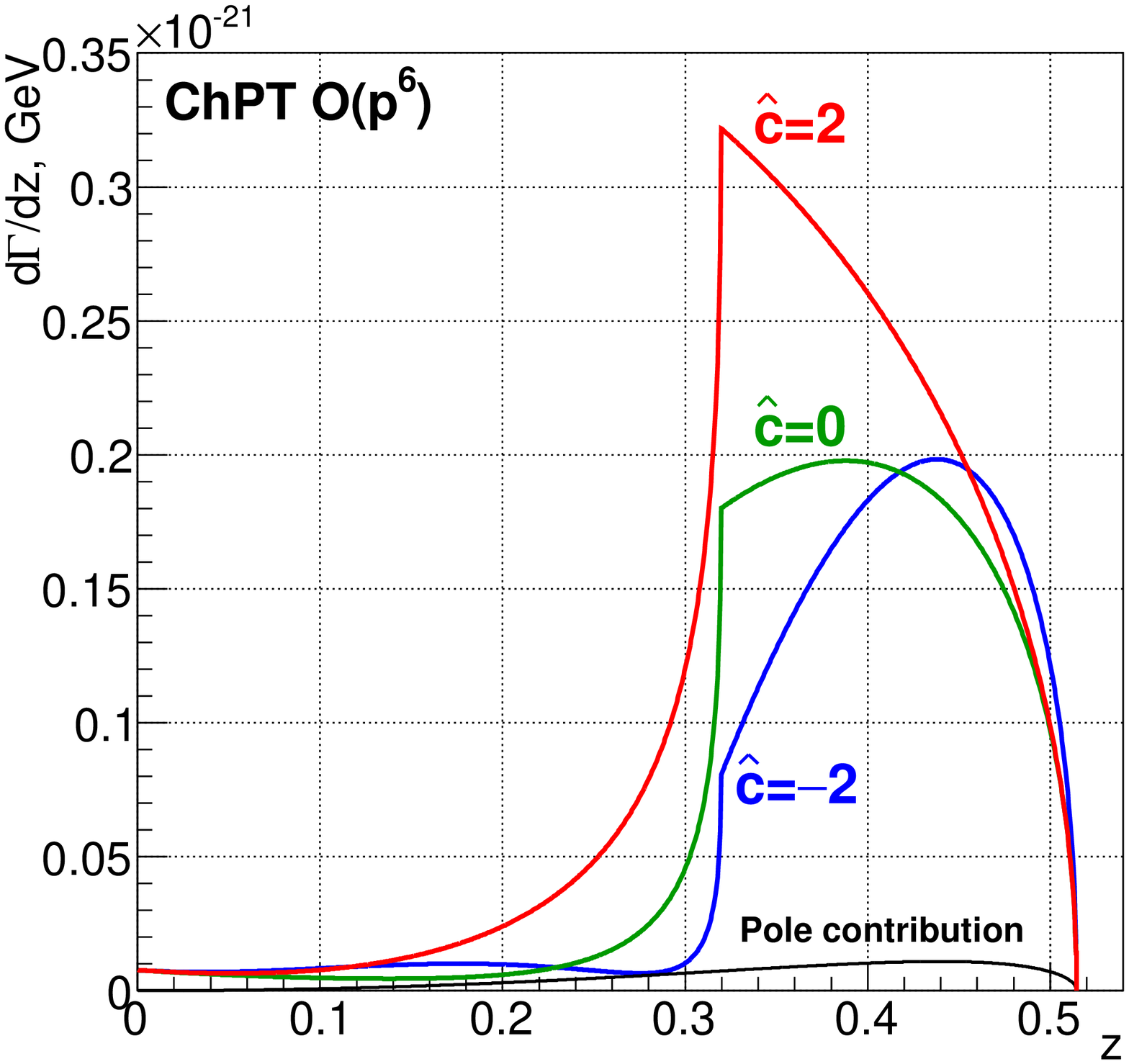}}
\end{center}
\vspace{-6mm} \caption{Differential rate ($d\Gamma/dz$) of the $K_{\pi\gamma\gamma}$ decays according to the ${\cal O}(p^4)$ and ${\cal O}(p^6)$ descriptions~\cite{da96} for several values of $\hat c$. The $\hat c$-independent pole contribution is also shown. For the ${\cal O}(p^6)$ parameterization, values of polynomial contributions~\cite{da96} $\eta_i=0$ and $K^\pm\to3\pi^\pm$ amplitude parameters from a fit to experimental data~\cite{bi03} are used.}
\label{fig:chpt}
\end{figure}

In the ChPT framework, the $K_{\pi\gamma\gamma}$ decay receives two non-interfering contributions at lowest non-trivial order ${\cal O}(p^4)$: the pion and kaon loop amplitude depending on an a priori unknown ${\cal O}(1)$ parameter $\hat{c}$, and the pole amplitude. Higher order unitarity corrections from $K\to3\pi$ decays modify the decay spectrum significantly; in particular, they lead to non-zero differential decay rate at zero diphoton invariant mass~\cite{da96}. The total decay rate is predicted to be ${\cal B}(K_{\pi\gamma\gamma}) \sim 10^{-6}$, with the pole amplitude contributing 5\% or less~\cite{da96, ge05}. The ChPT predictions for the $z=(m_{\gamma\gamma}/m_K)^2$
spectrum, where $m_{\gamma\gamma}$ is the diphoton invariant mass, for several values of $\hat{c}$ are presented in Fig.~\ref{fig:chpt}. These spectra exhibit a characteristic cusp corresponding to twice the pion mass due to the dominant pion loop amplitude.

Experimentally, the only published $K_{\pi\gamma\gamma}$ observation is that of 31 $K^+$ decay candidates in the kinematic region $100~{\rm MeV}/c<p_\pi^*<180~{\rm MeV}/c$ ($p_\pi^*$ is the $\pi^+$ momentum in the $K^+$ frame) by the BNL E787 experiment~\cite{ki97}.

\boldmath
\section{Measurements of the $K^\pm\to\pi^\pm\gamma\gamma$ decay}
\unboldmath

New measurements of the $K_{\pi\gamma\gamma}$ decay have been performed using two minimum bias data sets:  1) two special $K^\pm$ decay samples collected by the NA48/2 experiment at $\sim 10\%$ of the nominal beam intensity during 12 hours in 2003 and 54 hours in 2004; 2) a subset of the NA62-$R_K$ data sample collected over the whole duration of the data taking with downscaled trigger conditions with an effective downscaling factor of about 20. The employed trigger conditions required a time coincidence of signals in both HOD planes within the same quadrant and an energy deposit of at least 10~GeV in the LKr calorimeter. The resulting effective kaon fluxes used for the NA48/2 and NA62-$R_K$ $K_{\pi\gamma\gamma}$ analyses are similar, but the background conditions and resolution on kinematic variables differ significantly. The $K_{\pi\gamma\gamma}$ decay rate was measured with respect to the normalization decay chain with a large and well known branching fraction: the $K^\pm\to\pi^\pm\pi^0$ decay followed by the $\pi^0\to\gamma\gamma$ decay. Signal and normalization samples have been collected with the same trigger logic.

\begin{figure}[tb]
\begin{center}
\vspace{-2mm}
\resizebox{0.45\textwidth}{!}{\includegraphics{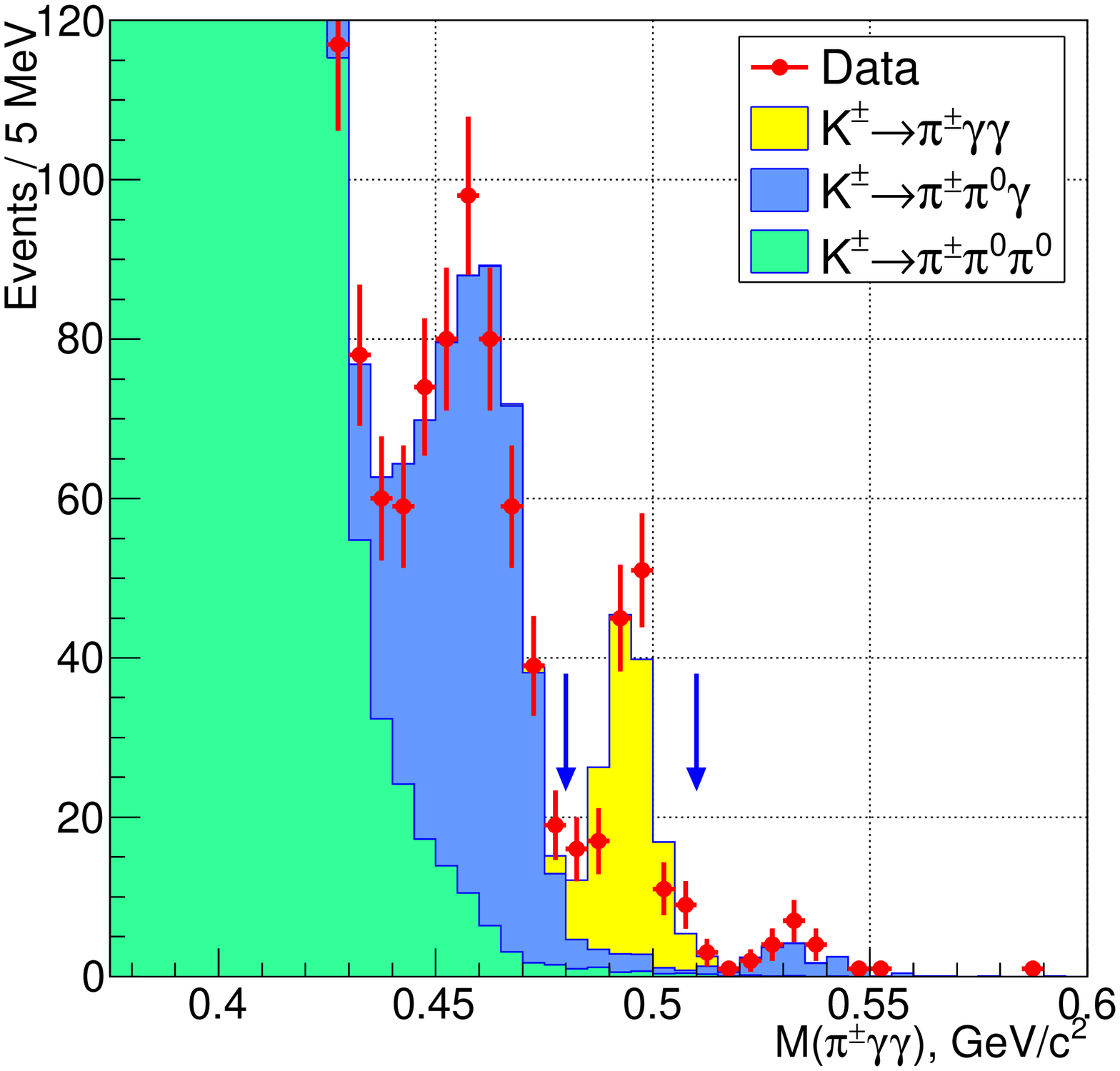}}%
\resizebox{0.45\textwidth}{!}{\includegraphics{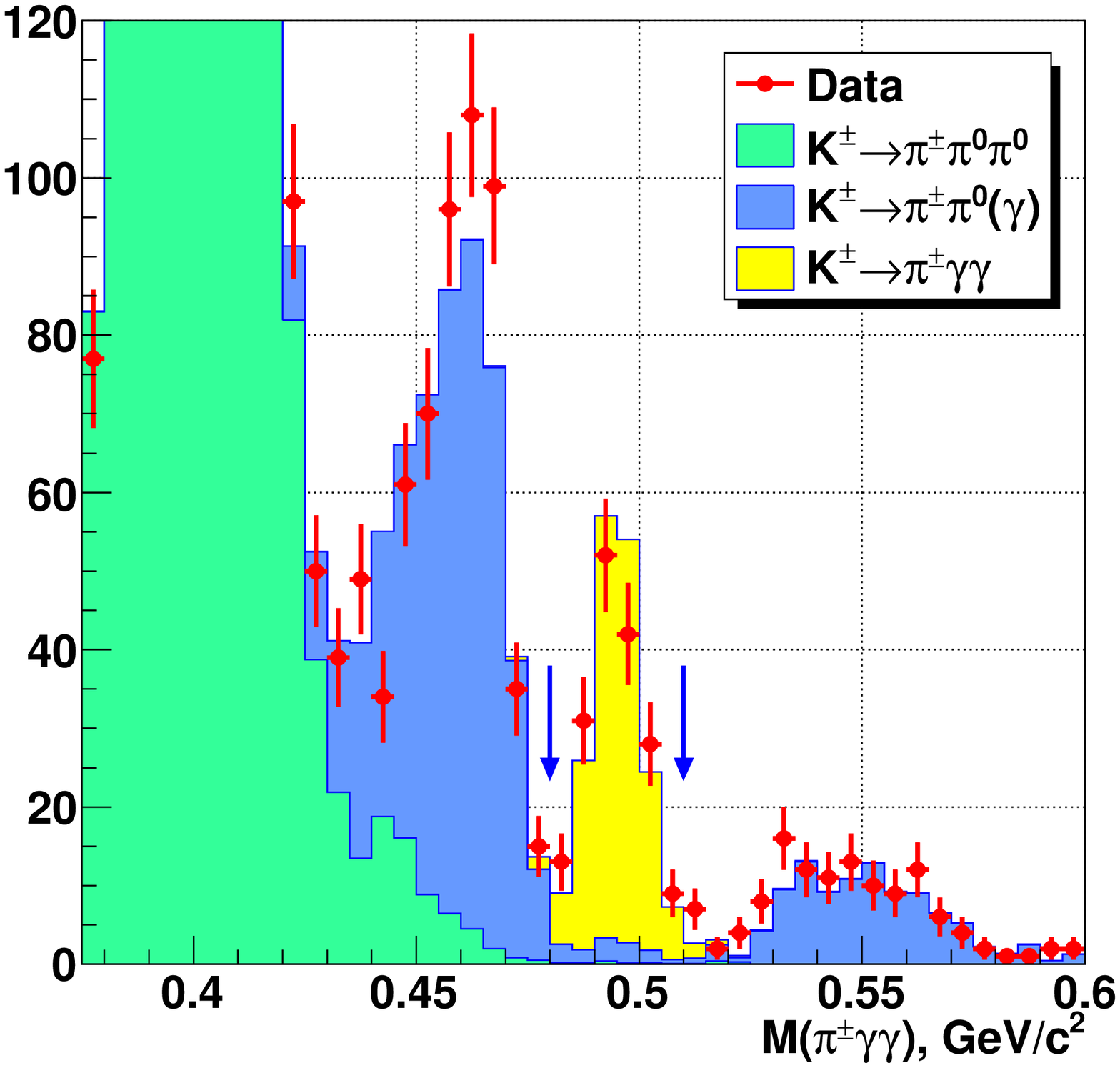}}%
\end{center}
\vspace{-9mm} \caption{The spectra of $\pi^\pm\gamma\gamma$ invariant mass with MC expectations for signal and backgrounds: NA48/2 data (left) and NA62-$R_K$ data (right). The signal region is indicated with arrows.} \vspace{-5mm} \label{fig:pigg-m}
\end{figure}

Signal events are selected on the basis of spectrometer and LKr calorimeter information in the kinematic region $z=(m_{\gamma\gamma}/m_K)^2>0.2$ to reject the $K^\pm\to\pi^\pm\pi^0$ background, as well as other backgrounds from the $\pi^0$ decays, peaking at $z=(m_{\pi^0}/m_K)^2=0.075$. The residual background contamination is due to $K^\pm\to\pi^\pm\pi^0\gamma$ and $K^\pm\to\pi^\pm\pi^0\pi^0$ decays, with photons either missing the LKr acceptance of forming merged clusters in the LKr calorimeter. The event selection includes an upper limit for the transverse size of the LKr clusters, which reduces the background due to merged clusters. The $\pi^\pm\gamma\gamma$ invariant mass spectra of the selected $K_{\pi\gamma\gamma}$ candidates, with the expectations of the signal and background contributions from MC simulations, are displayed in Fig.~\ref{fig:pigg-m}: 149 (175) decay candidates with a background contamination of 10\% (7\%) are observed in the NA48/2 (NA62-$R_K$) data set.

\begin{figure}[tb]
\begin{center}
\vspace{-2mm}
\resizebox{0.33\textwidth}{!}{\includegraphics{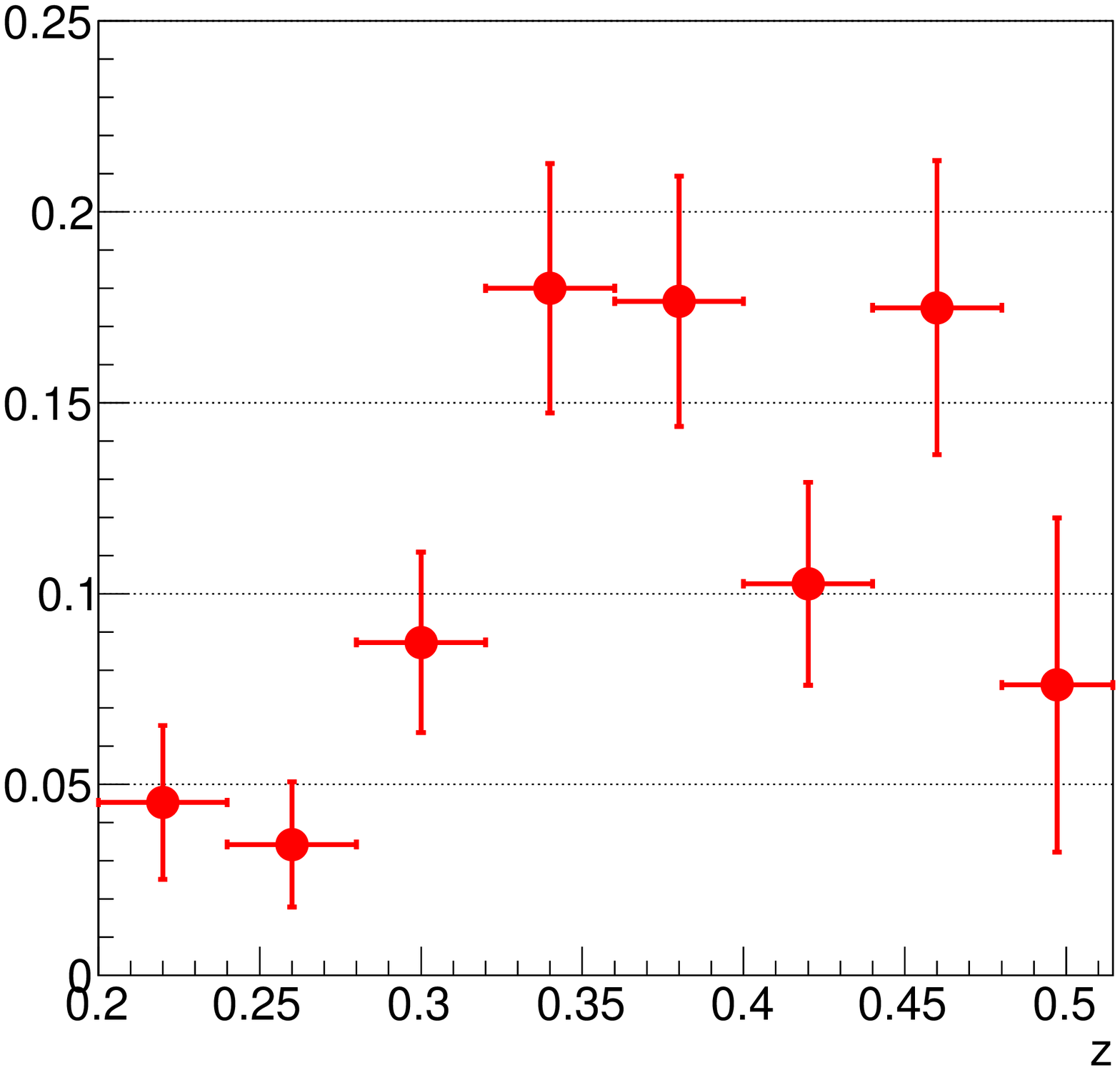}}%
\resizebox{0.33\textwidth}{!}{\includegraphics{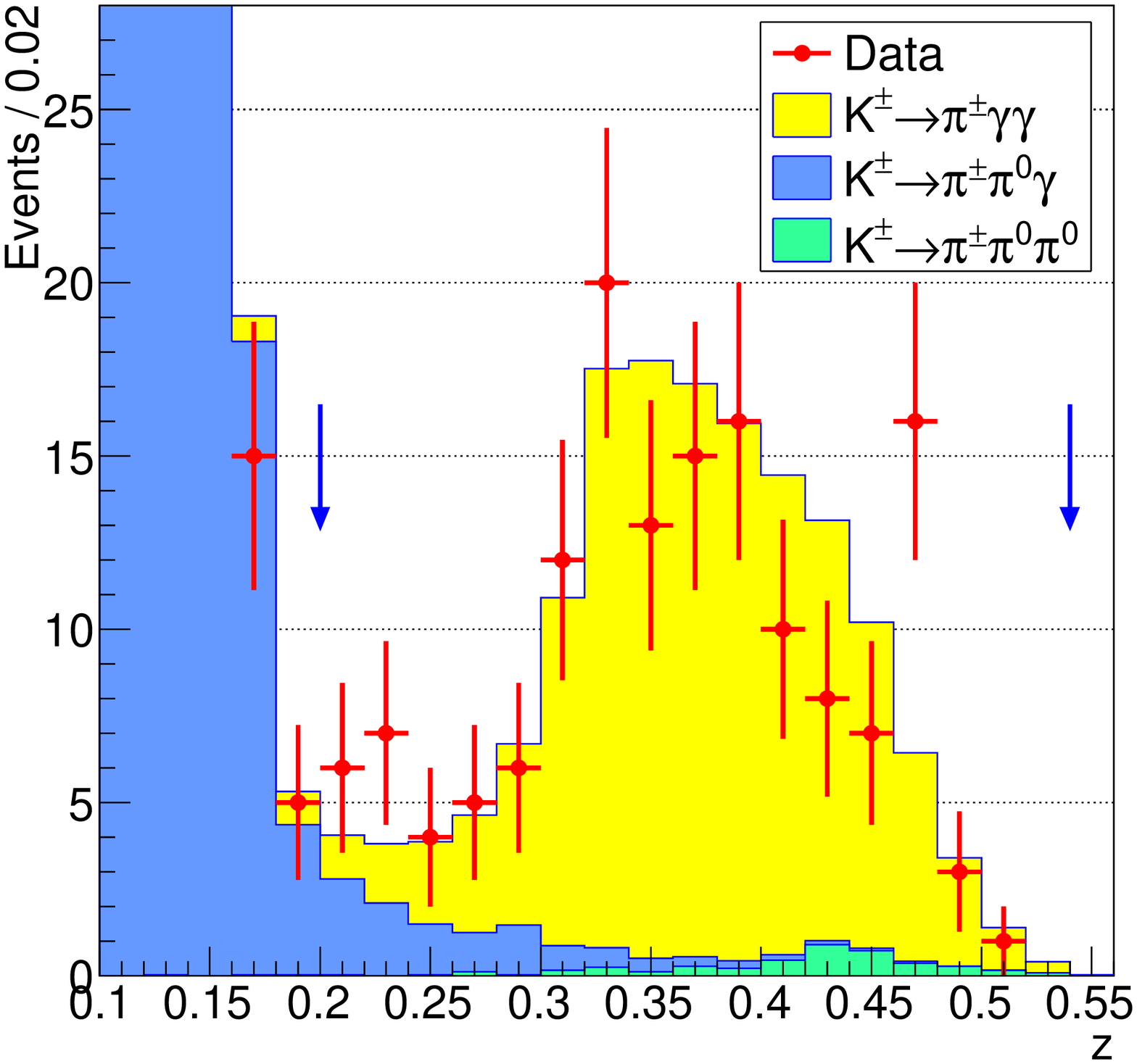}}%
\resizebox{0.33\textwidth}{!}{\includegraphics{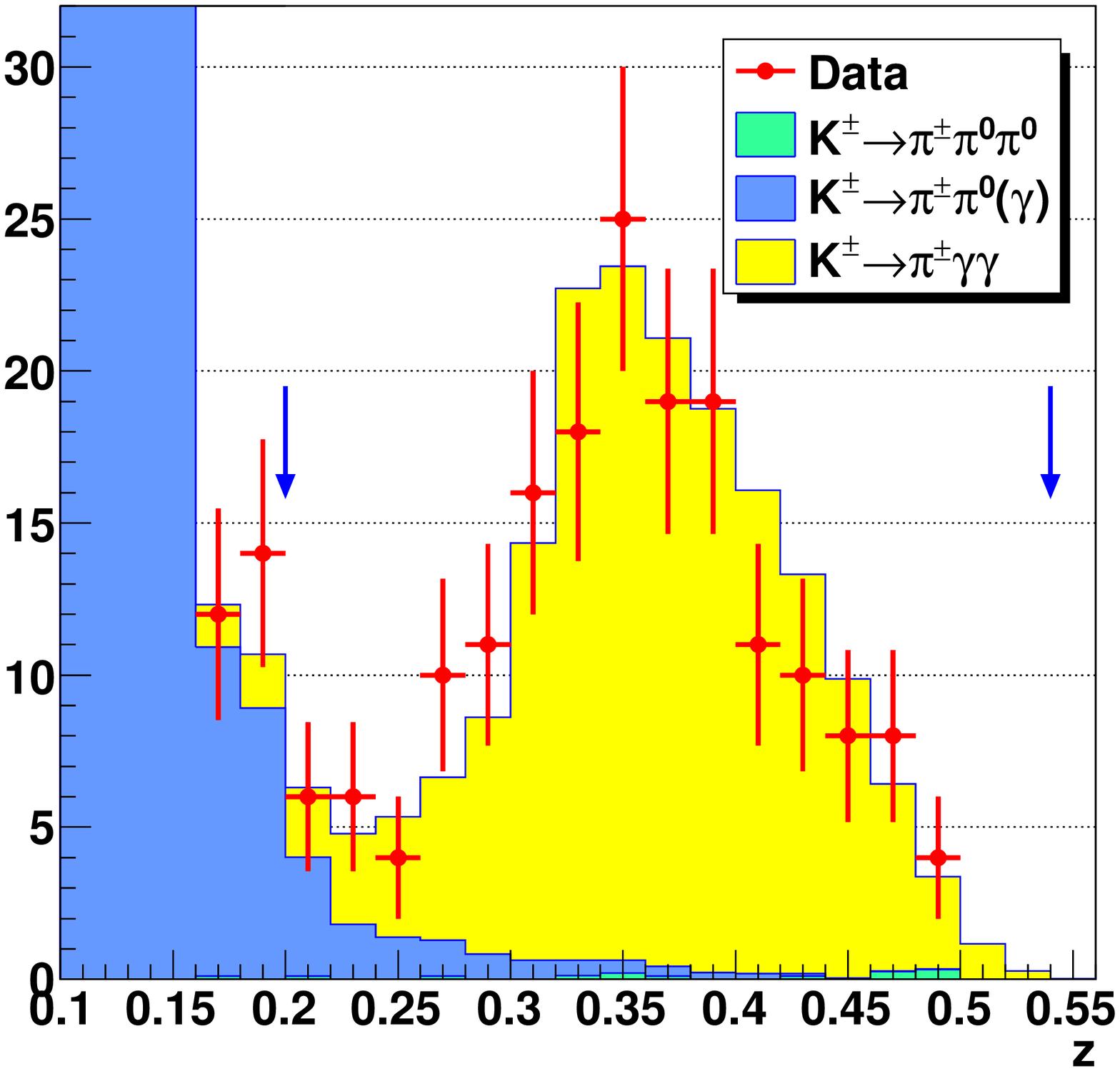}}%
\put(-434,138){\small ${\cal B}(K_{\pi\gamma\gamma})\times 10^6$}
\end{center}
\vspace{-8mm} \caption{Left: measurements of partial model-independent branching fractions of the $K_{\pi\gamma\gamma}$ decay in bins of the $z$ kinematic variable from the NA48/2 data sample. The spectra of $z=(m_{\gamma\gamma}/m_K)^2$ with MC expectations for signal (best fit) and backgrounds: NA48/2 data (centre) and NA62-$R_K$ data (right). The signal region ($0.2<z<0.52$) is indicated with arrows.} \vspace{-4mm} \label{fig:pigg-z}
\end{figure}

A model-independent measurement of the $z$ spectrum in the kinematic range $z>0.2$ has been performed for the NA48/2 data set. The partial branching fractions in bins of the $z$ variable have been measured: they are model-independent because the considered $z$ bin width is sufficiently small for the acceptances in $z$ to have a negligible dependence on the assumed $K_{\pi\gamma\gamma}$ kinematical distribution. In addition, the $y$-dependence of the differential decay rate expected within the ChPT framework is weak~\cite{da96,ge05}. The final results of the measurements of the partial branching fractions in bins of the $z$ variable are presented in Fig.~\ref{fig:pigg-z} (left). The model-independent branching fraction in the kinematic region $z>0.2$ is computed by summing over the $z$ bins: ${\cal B}_{\rm MI}(z>0.2) = (0.877 \pm 0.087_{\rm stat} \pm 0.017_{\rm syst}) \times 10^{-6}$.

Measurements of the ChPT parameter $\hat c$ have been made for both NA48/2 and NA62-$R_K$ data samples by performing log-likelihood fits to the reconstructed $z$ spectra. The data spectra of the $z$ kinematic variable, together with signal and background expectations from simulations, are displayed in Fig.~\ref{fig:pigg-z} (centre, right): they support the ChPT prediction of the cusp at two-pion threshold. The values of the $\hat{c}$ parameter in the framework of the ChPT ${\cal O}(p^4)$ and ${\cal O}(p^6)$ parameterizations according to the formulation~\cite{da96} have been measured by the performing likelihood fits to the data. The ${\cal O}(p^6)$ parametrization involves a number of external inputs. In this analysis, they have been fixed as follows: the polynomial contribution terms are $\eta_1=2.06$, $\eta_2=0.24$ and $\eta_3=-0.26$ as suggested in~\cite{da96}, while the $K^\pm\to3\pi^\pm$ amplitude parameters come from a fit to the experimental data~\cite{bi03}.

The preliminary results of the fits are presented in Table~\ref{tab:chat}: they are in agreement with the earlier BNL E787 ones. The uncertainties are dominated by the statistical ones; the systematic errors are mainly due to uncertainties of the background estimates. A combination of results from the two experiments has been performed, taking into account the large positive correlation of the systematic uncertainties of the two measurements. The combined results are also presented in Table~\ref{tab:chat}. The branching ratio in the full kinematic range corresponding to the combined value of the $\hat c$ parameter within the ${\cal O}(p^6)$ formulation is ${\cal B}(K_{\pi\gamma\gamma})=(1.01\pm0.06)\times 10^{-6}$.

\begin{table}[tb]
\begin{center}
\caption{The preliminary results of the fits to the $K^\pm\to\pi^\pm\gamma\gamma$ diphoton mass spectra to the ChPT parameterizations~\cite{da96}.}
\label{tab:chat}
\begin{tabular}{l|ccc}
\hline & NA48/2 measurement & NA62-$R_K$ measurement & Combined\\
\hline
$\hat{c}$, ${\cal O}(p^4)$ fit &
$1.36\pm0.33_{\rm stat}\pm0.07_{\rm syst}$ &
$1.71\pm0.29_{\rm stat}\pm0.06_{\rm syst}$ &
$1.56\pm0.22_{\rm stat}\pm0.07_{\rm syst}$\\
$\hat{c}$, ${\cal O}(p^6)$ fit &
$1.67\pm0.39_{\rm stat}\pm0.09_{\rm syst}$ &
$2.21\pm0.31_{\rm stat}\pm0.08_{\rm syst}$ &
$2.00\pm0.24_{\rm stat}\pm0.09_{\rm syst}$\\
\hline
\end{tabular}
\end{center}
\vspace{-2mm}
\end{table}

\end{document}